\documentclass[aps,twocolumn,prl,letter,showpacs,showkeys,preprintnumbers,amsmath,amssymb]{revtex4}

\usepackage{graphicx}
\usepackage{dcolumn}
\usepackage{bm}

\begin{document}


\title{
Synchronization and oscillator death in oscillatory media with stirring}

\author{Zolt\'an Neufeld$^1$, Istv\'an Z. Kiss$^2$, Changsong S. Zhou$^3$ and J\"urgen Kurths$^3$}

\affiliation
{
$^1$ Center for Nonlinear Studies,
Los Alamos National Laboratory, New Mexico 87545 USA
}

\affiliation
{
$^2$Department of Applied Mathematics,
University of Leeds, LS2 9JT Leeds, United Kingdom
}

\affiliation
{
$^3$Institute of Physics, University of Potsdam
PF 601553, 14415 Potsdam,  Germany
}

\date{\today}

\begin{abstract}
The effect of stirring in an inhomogeneous oscillatory medium is investigated. 
We show that the stirring rate can control the macroscopic behavior 
of the system producing collective oscillations (synchronization) 
or complete quenching of the oscillations (oscillator death). 
We interpret the homogenization rate due to mixing as a measure
of global coupling and compare the phase diagrams
of stirred oscillatory media and of populations of globally coupled
oscillators.
\end{abstract}

\pacs{PACS}
\maketitle

The problem of synchronization of a large population of non--linear
oscillators has received a great deal of attention lately \cite{sync} 
due to its applications in a variety of physical \cite{josephson}, chemical 
\cite{chem} and biological
systems \cite{winfree,yeast}, social phenomena \cite{clapp} etc.
Typically two types of couplings are used to describe the interactions
in such systems:
{\it i.) global coupling},
where each of the oscillators is coupled to all the others,
or {\it ii.) local
coupling}, where only the nearest neighbors are interacting.
Global coupling is relevant for oscillators
communicating via visual or acoustic signals, like flashing fireflies, chirping
crickets, clapping audiences, and can also be implemented 
by electric coupling (see \cite{chem} for a recent experiment on synchronization
of globally coupled electrochemical oscillators). In these cases
the time necessary for information to spread over the whole
system is much shorter than the period of the oscillations.
In general, global coupling leads to synchronization when the coupling
is sufficiently strong and the distribution of the natural frequencies
is not too broad.

In a continuous oscillatory medium (e.g. reaction-diffusion system)
the oscillations at different points of the medium interact through
molecular diffusion. Since the time-scale of diffusive transport on macroscopic 
lengthscales ($L$) is typically much longer than the characteristic
timescale of the oscillations, $L^2/D \gg T_{osc}$,
diffusion is unable to produce synchronized oscillations over the whole
domain and the only coherent behavior appears in form of propagating 
waves \cite{wave1,wave2}.

In certain situations the oscillators are embedded into a moving medium,
e.g. in a fluid flow. Examples are 
oscillatory chemical or biological systems in stirred reactors 
(Belousov-Zhabotinsky reaction \cite{wave2}, metabolic 
oscillations in cell suspensions \cite{yeast}), or in geophysical
context: oceanic plankton populations \cite{plankton} and chemical 
reactions in the atmosphere \cite{atmos} transported by 
large scale geophysical flows.
However, the effect of stirring in oscillatory media has not yet 
been investigated. 
It is often assumed, that strong stirring
leads to spatially uniform concentrations, and thus the temporal evolution,
simply described by a set of ordinary differential equations, becomes
independent of the stirring process.
But in most real systems there are inherent inhomogeneities
imposed by boundary conditions or non-uniformities of certain 
external parameters. This can be due to
spatial variations of temperature or illumination in a chemical reactor,
or non-uniform distribution of sources in environmental flows.
Therefore perfectly uniform concentrations 
are unattainable and stirring effects cannot be ignored. 

In this Letter we investigate the behavior of a stirred
oscillatory medium, 
described by a set of reaction-advection-diffusion equations
\begin{equation}
{\partial_t C_{i}} + {\bf v}({\bf r},t)\nabla
C_{i}=F_{i}(C_{1},...,C_{N};{\bf r})+D \Delta C_{i},
\label{rad}
\end{equation}
where $C_{i}({\bf r},t), i=1,2,...,N$, are the concentrations
of $N$ interacting species advected by an incompressible
fluid flow ${\bf v} ({\bf r},t)$.
The velocity field is assumed to be time-dependent,
that ensures efficient mixing by the chaotic motion of the fluid 
elements \cite{mixing}. 
The functions $F_{i}$ describe the
nonlinear interactions between the components (chemical reactions,
evolution of biological populations etc.) such that the local dynamics
has a stable limit cycle in each point of the medium
\begin{equation}
\dot C_i = F_{i}(C_{1},...,C_{N};{\bf r}) \to C_i(t+T_{\bf r};{\bf r})=C_i(t;{\bf r}).
\end{equation}
The explicit dependence of the interaction terms on the
spatial coordinate accounts for inhomogeneities of
the medium.
We consider the simplest form of inhomogeneity,
when 
the medium is composed by identical oscillators
except for their frequencies, that is non-uniform in space
\begin{equation}
F_{i}(C_{1},...,C_{N};{\bf r}) = (1+ \delta f({\bf r}))F'_{i}(C_{1},...,C_{N}),
\end{equation}
where $f({\bf r})$ and $\delta$ describe the shape
and the amplitude of the inhomogeneity. We assume, for 
simplicity, that $\langle f({\bf r}) \rangle = 0$ and 
$\langle f^2({\bf r}) \rangle =1$, where $\langle \;\; \rangle$ 
represents averaging over the domain. 

\begin{figure}
{\centering \resizebox*{0.8\columnwidth}{!}{\includegraphics{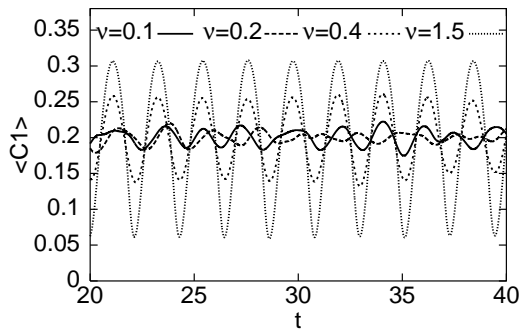}}}
\caption
{Time-dependence of the mean concentration $\langle C_1 \rangle$
for different stirring rates ($\nu$) and $\delta =0.2$.}
\end{figure}

In the numerical simulations stirring is modelled by a
sinusoidal shear flow with
alternating directions
\begin{eqnarray}
v_x = (A/T) \sin (2\pi y + \phi_n), v_y = 0 \nonumber \\
\;\;\; \;\;\;\; {\rm for} \;\;\;\; t \in [nT,(n+1/2)T) \nonumber \\
v_x = 0, v_y = (A/T) \sin (2\pi x + \phi_n) \nonumber\\ 
\;\;\; \;\;\;\; {\rm for} \;\;\;\; t \in [(n+1/2)T,(n+1)T)
\end{eqnarray}
($A=1.4$) advecting chaotically all fluid particles within 
the unit square
with doubly periodic boundary conditions.
The phases $\phi_n$ are chosen at random in each half
period. This ensures that there are no transport barriers
and all fluid elements can approach any other fluid
element in the domain.
We note that chaotic advection is a generic feature of simple time-dependent 
flows, therefore the results are expected to be characteristic to 
a broad range of fluid flows. 
We define the stirring rate as $\nu \equiv 1/T$, that is controlled
by the period of the flow.

For the oscillatory dynamics the well known
Lengyel-Epstein model of the chlorine--iodine--malonic acid
reaction (CDIMA) \cite{wave2} is considered
\begin{eqnarray}
F'_1(C_1,C_2)=1-C_{1}-{4C_{1}C_{2}}/({b +C_{1}^{2}}) \nonumber \\
F'_2(C_1,C_2)=a(C_{1}-{C_{1}C_{2}}/({b +C_{1}^{2}})).
\end{eqnarray}
The chemical dynamics has a uniform steady state,
($C_{1}^{*}=0.2,C_{2}^{*}=b+0.04$) which is unstable 
for the parameter values used, $b=0.005$, $a=0.375$,
and the only attractor is a limit cycle.
The shape of the inhomogeneity is chosen to be $f(x,y)=\sqrt 2 \sin(2\pi(x-y))$.
The system (3-5) is investigated for different
stirring rates $\nu$ and degrees of inhomogeneity $\delta$.
($D=10^{-4}$ in all simulations.)

Let us first discuss some special cases.
For non-reactive components ($F(C_1,..,C_N)\equiv 0$) an initially non-uniform
concentration is homogenized by mixing.
In flows with
chaotic advection the decay of the spatial fluctuations is exponential in time
\cite{eigenmode}
\begin{equation}
\langle (C-\langle C\rangle )^2\rangle \sim \exp(-2\alpha t),
\end{equation}
and for long times the spatial structure is dominated by the eigenmode of the
advection-diffusion operator ${\cal L}_{mix} \equiv D\Delta-{\bf v} \cdot \nabla$
with the largest (least negative) eigenvalue. 
\begin{equation}
C({\bf r},t) \to \langle C({\bf r},0) \rangle + e^{-\alpha t} \Phi_1, \;\;\;\; {\cal L}_{mix} \Phi_i = -\alpha_i \Phi_i.
\end{equation}
Strictly speaking the eigenmodes have a temporal dependence following
the evolution of the velocity field, but they are stationary at least in a
statistical sense.

In a uniform oscillatory medium, $\delta =0$,
the advection-reaction-diffusion problem has a spatially uniform
oscillatory solution. This is
stable to spatially non-uniform perturbations,
and numerical simulations suggest that it is globally
attracting.
The decay of the spatial
fluctuations is controlled by the chaotic mixing, as indicated by the 
exponential decay of the variance with the same exponent as for the 
non-reactive case. The only difference is that there are oscillations
superposed due to the oscillatory nature of the chemical dynamics, i.e.
$\langle (C-\langle C\rangle )^2\rangle \sim h(t) \exp(-2\alpha t)$ where 
$h(t)$ is periodic with the period of the oscillations of the mean field.

\begin{figure}
{\centering \resizebox*{0.8\columnwidth}{!}{\includegraphics{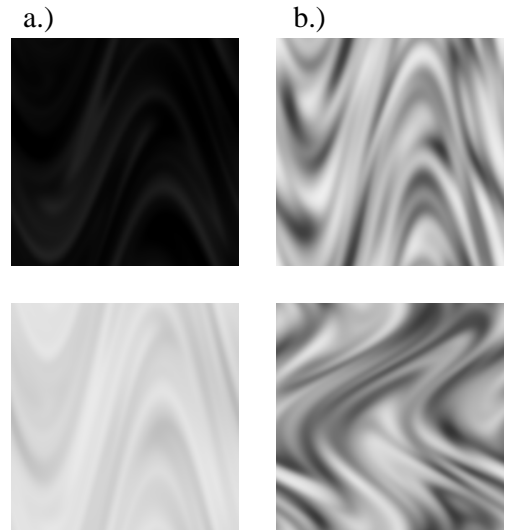}}}
\caption
{Snapshots of the concentration field $C_1(x,y)$ in the
synchronized ($a.) \nu = 0.4$) and the non-synchronized ($b.) \nu =0.167$)
regimes ($\delta = 0.2$)}
\end{figure}

Let us now consider 
a fixed amplitude of the inhomogeneity and vary the stirring rate.
When stirring is strong 
the mean concentration has an oscillatory time-dependence 
indicating synchronization (Fig.1),
but unlike in the case of the homogeneous medium,
the spatial fluctuations do not disappear completely, since
there is no spatially uniform oscillatory solution for $\delta>0$. 
For very fast stirring the spatial fluctuations
are weak and the oscillations of the mean concentrations 
are almost the same as for a uniform medium.
The amplitude of the
oscillations of the mean field decreases with the stirring rate, 
while the frequency remains the same.
On longer time scales, there is also a
weak irregular modulation of the amplitude, that becomes more pronounced for
slow stirring. 
When the stirring rate falls below a certain critical value,
the synchronized 
oscillations disappear, and the mean concentration is almost constant
apart from small irregular fluctuations.  

\begin{figure}
{\centering \resizebox*{0.8\columnwidth}{!}{\includegraphics{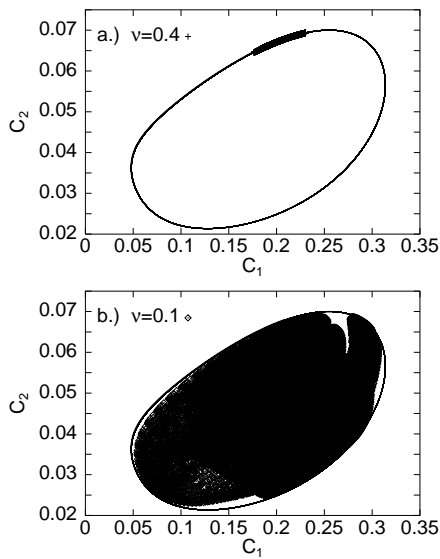}}}
\caption
{Projection of the concentration field $C_1(x,y)$ on the chemical
phase plane $C_1-C_2$ in the synchronized (a.) and non-synchronized case,
$\delta=0.2$. The limit cycle for a single
oscillator is also shown for reference.}
\end{figure}

Two pairs of snapshots of the concentration field $C_1(x,y)$, 
for the synchronized (a.) and unsynchronized (b.) case, 
are shown in Fig.2.
The complex spatial structure, characteristic to chaotic mixing, 
is combined with a coherent time-evolution 
for supercritical stirring,
while in the slow stirring case the snapshots corresponding to 
different times are statistically equivalent.
The difference between the two regimes is also clearly visible
on the projections of the concentration field onto the chemical
phaseplane $C_1-C_2$ (Fig.3). When the stirring is strong the projected
concentration fields appear as a small clump moving around the limit 
cycle of a single oscillator, while in the unsynchronized regime they 
extend over a large domain, that remains almost
unchanged in time. 

To characterize the degree of synchronization we calculate the
standard deviation in time of the spatially averaged concentrations
\begin{equation}
R \equiv \sqrt{ \frac{1}{\tau} \int_t^{t+\tau} \langle C \rangle^2 dt' - \left (\frac{1}{\tau} \int_t^{t+\tau}
\langle C \rangle dt' \right )^2},
\end{equation}
for different stirring rates (Fig.4). Large values of $R$ indicate 
synchronization. $R(\nu,\delta)$ can be normalized 
by dividing it with the same quantity obtained for the uniform 
oscillating medium, $R_0=R(\delta=0)$ ($0 < R/R_0 < 1$).
The order parameter $R/R_0$ 
increases
sharply above a critical stirring rate. In the unsynchronized regime,
below the critical stirring rate, the order parameter is small. 
We believe, that in this regime $R$ tends to zero in the $D \to 0$
limit, that is analogous to the limit of infinitely many oscillators
in the case of global coupling.

\begin{figure}
{\centering \resizebox*{0.75\columnwidth}{!}{\includegraphics{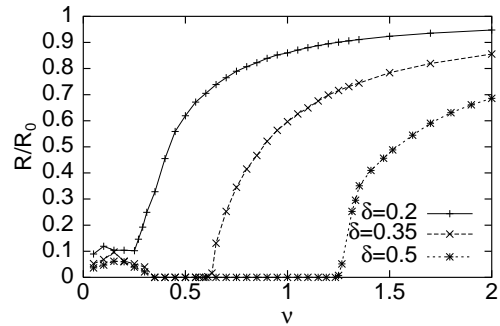}}}
\caption
{The normalized order parameter as a function of stirring rate
for different degrees of non-uniformity of the medium.}
\end{figure}

The critical stirring rate depends on $\delta$, faster stirring is needed 
for synchronization when the medium has a larger spread of the local
frequencies. When $\delta$ is sufficiently large
a new regime appears between the synchronized and unsynchronized states. 
At intermediate stirring rates the oscillations of the mean field disappear completely (R=0) and the concentrations become uniform in space.
This 'oscillator death' state corresponds
to the unstable equilibrium of the homogeneous chemical system, $C_i^*$. 
Thus, stirring in the presence of inhomogeneity can stabilize the
unstable steady state of the chemical dynamics and suppress the
oscillations. This results from the competition between the inhomogeneity
of the medium generating non-uniform concentrations and mixing that tends
to reduce the spatial fluctuations.

Similar regimes 
have been observed in ensembles of globally coupled oscillators 
\cite{Ermentrout,PRL}.
In the stirred media 
due to the relative displacement of different parts of
the medium the neighborhood of each point is changing in time,
thus regions that are initially far from each other
can interact at a later time. This defines a characteristic timescale
of mixing as the time needed to bring pairs of points, initially
separated by a distance comparable with the size of the domain, sufficiently
close so that they can interact by diffusion.
When this time is much shorter than the period of the oscillations,
there is an effective global interaction in the moving medium.
Thus the character and strength of the coupling can be controlled by the
stirring rate.
An analogous situation occurs in an ensemble of
oscillators coupled through a network in which each node has a small number
of connections that are changing in time in a random fashion \cite{network}.

\begin{figure}
{\centering \resizebox*{0.75\columnwidth}{!}{\includegraphics{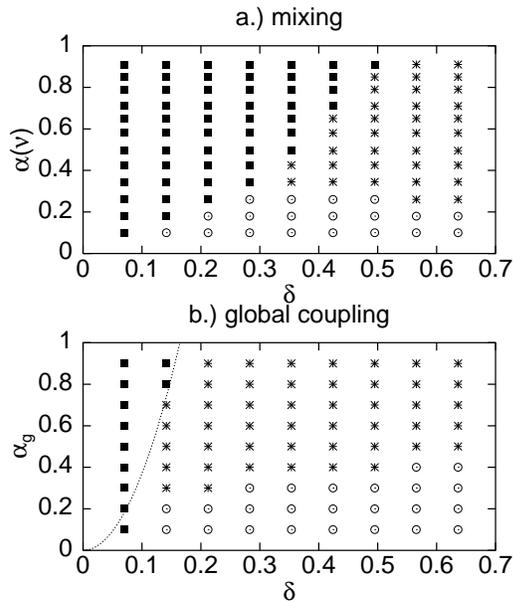}}}
\caption
{Phase diagram of the oscillatory medium with stirring (a.)
and global coupling (b.) - the symbols represent: synchronization
($\blacksquare$), oscillator death ($\ast$) and no synchronization
($\odot$). The dashed line is the phase boundary based on equation (10).}
\end{figure}

The dynamics of globally coupled oscillators in the continuum limit is described by
\begin{equation}
\partial_t C_i = (1+\delta f({\bf r}))F_i(C_1,..,C_N) + \alpha_g (\langle C_i \rangle - C_i).
\end{equation}
Similarly to the mixed system,
in the absence of oscillations ($F_i(C_1,..,C_N)\equiv 0$),
global coupling leads 
to an exponential decay of the concentration fluctuations, 
${\rm Var}(C) \sim \exp(-2 \alpha_g t)$.
Based on this analogy, we interpret the homogenization rate 
in the stirred system, $\alpha$, as a measure of the coupling 
strength resulting from the combined effects of advection and diffusion.

In Fig.5 we present phase diagrams, 
both for the case
of mixing and global coupling. The homogenization rate, $\alpha$, corresponding
to different stirring rates have been obtained numerically by measuring
the decay rate of the variance of the concentration field for
a non-reactive component. In the fast stirring  limit ($\nu L^2/D \gg 1$)
the homogenization rate tends to be proportional with the stirring rate.
The two phase diagrams have a qualitatively similar structure.
Both strong coupling and fast stirring leads to synchronization, 
or oscillator death when the inhomogeneity of the medium is strong.
Slow stirring is analogous to weak coupling as shown by the lack
of synchronization in this regime.

For globally coupled oscillators an approximation for the 
boundary between the synchronization and oscillator death phase has been found 
recently in \cite{PRL}  
\begin{equation}
\alpha_c(\delta) = \lambda \left [ \left ({ \omega}/{ \lambda } \right )^2 - 1 \right ] \delta^2,
\label{bound} 
\end{equation}
where $\lambda$ and $\omega$ are the real and complex parts of 
the eigenvalue of the unstable steady state $C^*$.
The above result is valid when $\omega \gg \lambda$,
in our case $\lambda=0.222$ and $\omega=2.878$. We
find that in the case of global coupling the boundary (\ref{bound}) agrees 
well with the numerical results. For the mixed system, however, the
same boundary is shifted toward lower stirring rates/stronger inhomogeneity.

This is not surprising since the advection-diffusion operator has a
complex structure and is not simply equivalent to a linear relaxation 
to the mean concentration. 
Although the long time decay of the fluctuations is dominated by the eigenmode
with the largest eigenvalue, $\alpha=\alpha_1$, by replacing mixing with 
global coupling corresponding to the most slowly decaying eigenmode we 
underestimate the strength of the coupling due to mixing. This may be the 
explanation for the difference between the two diagrams in Fig.5.
Another origin of the deviation could be that in the system with mixing
fluid parcels move during a period of the flow and therefore the average
value of the shape function $f(x,y)$ calculated along fluid trajectories
has a smaller variance than in a motionless medium, resulting in a weaker
effective inhomogeneity of the medium.

In summary, we have shown that oscillatory
media with stirring exhibit qualitatively similar behavior to populations
of globally coupled oscillators, and the effective coupling strength is
controlled by the stirring rate. Changes in the stirring rate
can lead to transitions to synchronization or oscillator death.
This may explain some of the stirring effects observed in laboratory 
experiments \cite{Noszticzius} and could also be exploited for controlling the dynamics of 
oscillatory systems. Similar behavior may also arise in
systems where the local dynamics is chaotic.

We thank Silvia De Monte, Peter H. Haynes and Francesco d'Ovidio for useful 
discussions.






\end{document}